\documentclass[tighten, times, twocolumn]{aastex631}  
\usepackage{graphicx}
\usepackage{color}
\newcommand{\etal}{\mbox{\rm{et al.}~~}}
\newcommand{\kms}{\mbox{km\,s$^{-1}$}}
\begin{document}
\par
\vspace{-1cm}
\par
\title{Pregalactic globular cluster formation}

\author{Jeremy Mould}
\affiliation{Swinburne University}
\affiliation{ARC Centre of Excellence for Dark Matter Particle Physics}
\author{Jarrod Hurley$^1$}

\begin{abstract}
The QCD phase transition in the early universe may provide 
primordial black hole nuclei for globular clusters. We consider the accretion and star formation that follow, once 1000 M$_\odot$ nuclei have formed. When such a nucleus has formed, it remains. Whether these are common in the oldest globular clusters is one decidedly challenging question for the model, which is, as yet, unanswered; another is a possible contribution to the cosmic gravitational radiation background.

\end{abstract}

\section{Introduction}
The formation of globular clusters (GCs) has been a puzzle for astronomers ever since they were discovered as naked eye objects that could be resolved into stars with telescopes. 
They have been considered the oldest stellar objects in the Universe since the times of Baade (1944) and the first hydrogen burning ages of Haselgrove \& Hoyle (1956) and Sandage (1953). The first suggestion of a pregalactic origin was by Peebles \& Dicke (1968). Fall \& Rees (1985) argued that GCs form in the collapsing gas of a protogalaxy. Ricotti (2002) showed that GC formation may have played an important role in reionizing the intergalactic medium (IGM). Simulations by Moore \etal (2006) found that GCs appeared at  
z $ >$ 12, as the gas within protogalactic halos reached virial temperatures of 10$^4$K, cooled rapidly, and fragmented, while the models of Padoan, Jimenez \& Jones (1997) associated formation with H$_2$ cooling in 10$^8$ M$_\odot$ clouds.  Cen (2001) pointed to shocks caused by reionization, and Trenti,  Padoan \& Jimenez  (2015) to shock heating caused by halo mergers.

These milestone models and others
have been reviewed by Kruijssen (2025). A number of possibilities have been directed towards high redshift (Renzini 2017; Forbes \etal 2018, Trenti, Padova \& Jimenez 2015)
Their formation in the Magellanic Clouds in the last billion years, then, is a 
different process, unless the availability of unprocessed neutral hydrogen is a common factor, dwarf galaxies being the slowest to commence the initial burst of star formation. 
Recent starbursts have been clearly implicated in GC formation in work by Renaud, Bournaud \& Duc (2015) and Wilson \etal (2006), for example.

What makes primordial black holes (PBHs) attractive as the key instigator of star formation in both cases is their scale free availability over a broad range of masses with a log m + log n = constant initial mass function (IMF).  Whether PBH exist or not remains an open question, 
and the IMF is a matter of debate. According to Chen \& Hall (2024) monochromatically (i.e. single mass) PBH make up less than 10\% of the dark matter from 10$^{-5}$ to 1 M$_\odot$  and less than 1\% from 1 to 10$^3$ M$_\odot$. These constraints would be weakened by an order of magnitude for the IMF we assume here. 

\begin{figure*}	
\vspace{-1 cm}
\centering
\includegraphics[width=\textwidth,angle=0]{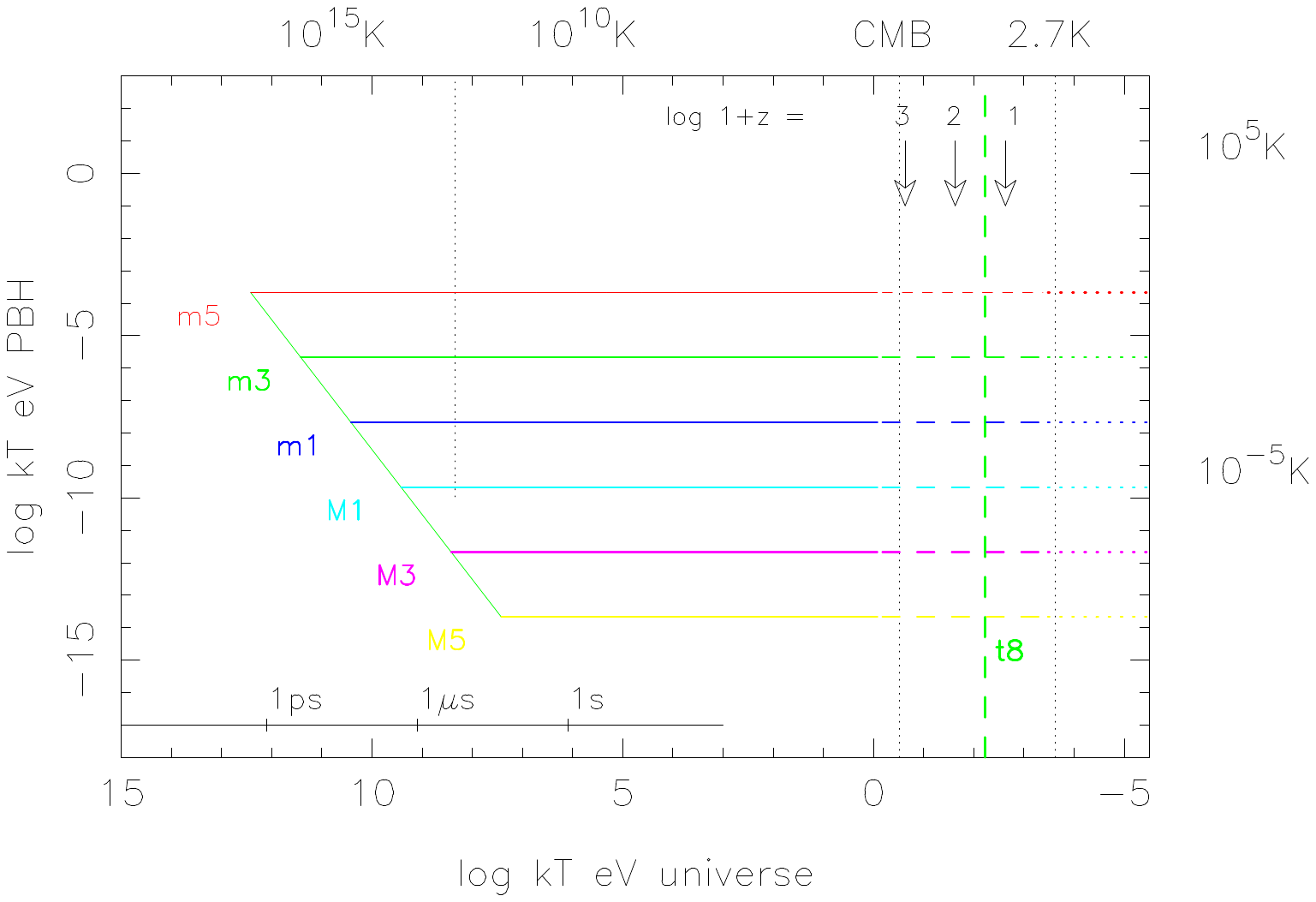}
\caption{Evolutionary tracks of PBHs. The horizontal axis is the temperature of the Universe and the vertical axis the Hawking temperature. The tracks become dashed lines as they pass into the matter dominated era and dotted in the dark energy dominated time. The vertical dotted lines are the present day (2.7K) and the QCD phase transition at 220 MeV, which intersects the green diagonal PBH birth line at M3. Galaxy formation is around 10$^8$ years (t8). The nomenclature m5 and M5 for PBHs denotes 10$^{-5}$ and 10$^5$ M$_\odot$ respectively. A timeline is given in microseconds etc after the Big Bang. Details are given by Mould (2025).
}
\vspace{1cm}
\end{figure*}
Observationally, there is evidence of GCs 
existing at redshift 4
(Senchnya \etal 2024, Vanzella \etal 2022). A pregalactic origin has been proposed (Bird \etal 2013, Beasley \etal 2003). In principle, the first objects on the scene in pregalactic times were PBHs.
Subsolar mass PBHs may have been detected (Niikura \etal 2019) and supermassive PBHs are candidates
for the progenitors of quasars (Bicknell \& Henriksen\footnote{Also: Volonteri, Habouzit \& Colpi (2021); Davies,
 Miller \& Bellovary, (2011), Lupi \etal (2014); Ziparo, Gallerani \& Ferrara (2024); Sobrinho
 \& Augusto (2024).} 1979, Mould \& Batten 2025).

We present PBH simulations in $\S$2. These form mini-halos, i.e. potential wells into which gas can fall. We go on to outline cluster star formation in $\S$3 and the observational evidence for nuclei consisting of accreted PBHs in $\S$4.
We conclude ($\S5$) that a model with pregalactic halos is currently viable, but a decisive test is within reach.
\begin{figure}[h]	
\includegraphics[width=.6\textwidth]{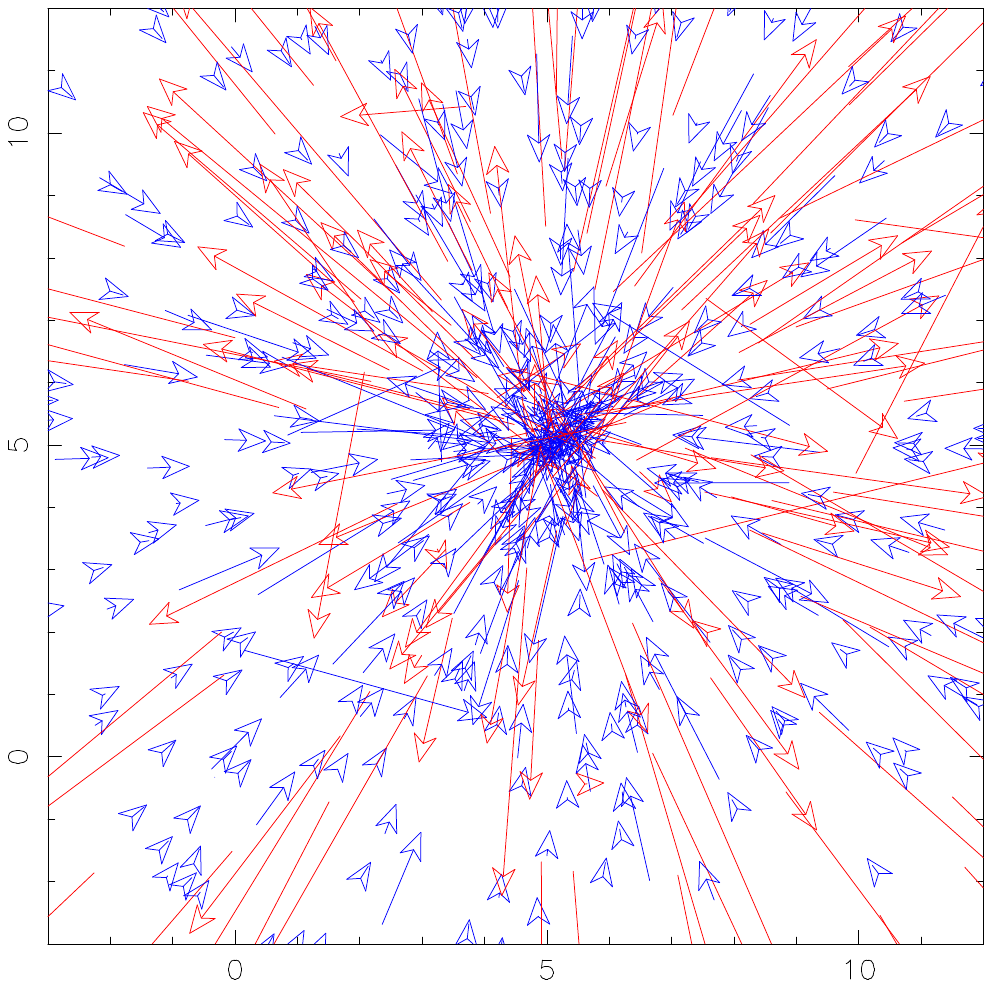}
\caption{Velocities of PBH particles towards the nucleus in blue, away in red. The scale is in pixels,
which are effectively AU.}
\end{figure}
\section{Accretion by 1000 M$_\odot$ PBH nuclei}
Carr (1981) was first to consider pregalactic accretion by black holes. At the time of matter radiation equality (z$_{eq} \approx$ 3400, Planck 2020) 
PBHs may already have formed (Figure~1) with an IMF spreading equal mass over each decade in mass (Mould 2025).

PBHs from 1 to 1000 M$_\odot$ 
are candidates  for PBH formation\footnote{A phase transition involves a large volume change at constant temperature. Compare the cosmic density at 220 MeV, a moderate 1.5~$\times$~10$^6$ gm cm$^{-3}$, comparable to a white dwarf,   with nuclear or pulsar density. In such volume changes, large density inhomogeneities are considered the most likely formation sites for PBHs} at the time of the QCD phase transition  (Alonso-Monsalve \& Kaiser 2023).   
PBH
mergers are considered by Carr \etal (2019). The mass distribution has a tail that reaches 100 M$_\odot$ (Musci, Jedamzik  \& Young 2023). Higher masses may be possible (Postnov \& Chekh 2024). Whether there is a further phase transition at the e$^+$e$^-$ annihilation transition, which might seed more massive PBHs, is an open question (Jedamzik 2025).  Carr \& Kuhnel’s (2021) 
review gives an equation for the characteristic mass of a PBH formed at the QCD transition, and Byrnes \etal (2018) provide the relevant mass distribution function. The maximum mass is 1000 M$_\odot$, and for dominance of its surroundings, that is what we use. The symbol M3 in Figure 1 marks their origin.  
Mould (2025) gives an equation for  PBH mass at a given temperature in the radiation dominated era, and for T$_{\rm{QCD}}$  $\sim$ 220 MeV this is $\sim$100 M$_\odot$. 

{}
The dominant mass in an overdensity will begin accreting its smaller companions.
We have studied this with a gravity-only n-body code and
 take the nuclear accretion rate from the Bondi-Hoyle (1944) formula and the sound speed from
Thomas, Kopp \& Skordis (2016). 
A typical velocity pattern of PBH particles near the nucleus is shown in Figure 2.

Before recombination, the gas would be fully coupled to the radiation and not to the dark matter. Even up until z $\sim$ 100, a order-unity fraction of the baryons would be coupled to the CMB radiation. In general the gas is significantly hotter than the dark matter and can also have a significant net relative velocity on small scales. We note again that there is no gas
in these simulations; they are dark-matter-only. But the dark matter forms a potential into which gas would be drawn.

\subsection{Simulation Results}

We made a number of 100000 particle n-body runs whose details are in Table 1 and illustrated in Figure 3. The run numbers are an internal reference code. 
The integration scheme was simply {\bf $\delta$v} = {\bf a}~dt, where {\bf a} is the acceleration, followed by {\bf $\delta$r} = {\bf v}~dt, where {\bf v} is the velocity. The numerical resolution is a minimum particle distance of 10$^{-20}$ 
imposed at the time the inter-particle distance, s, is calculated. Adaptive timesteps were
determined by requiring that v~dt $<<$ $<$s$>$, the mean value of s. The PBHs expand with the Hubble flow.
The initial conditions were a uniform random distribution of particles in a sphere with a nucleus of mass m, the largest in the distribution, and zero velocity. Other PBH particles range between m1 and m2.

The simulations are scale free in radius, but accretion is density dependent (dm/dt =
$\pi\sigma\rho v$), where $\sigma$ is the cross section, 
the square of the Bondi radius, $\rho$ the density
and $v$ the median 
velocity dispersion imparted by PBH inter-particle attraction.  Cosmological expansion scales the distances, not the velocities.
We adopt h = 0.73 and $\Omega_m$ = 0.3 to obtain $\rho$ in this equation, where the former is the dimensionless Hubble Constant and the latter the overall  matter density. The initial nuclear mass is in column (2), and the final mass
in the right hand column. Only the nucleus accretes. The IMF was a top hat
in log m with lowest and highest mass shown as m1 and m2.
Mould \& Batten (2025) have shown that that IMF in the case of supermassive
black holes yields the observed QSO luminosity function.
\begin{table}
	\caption{n-body simulations}
\begin{tabular}{lrrrrr}
\hline
	Run&	Initial m&m1&m2&Final m\\
	\#&nucleus&~~~~M$_\odot$&&nucleus\\
    \hline
	75*&1000&1&1000&1027&\\
	76&1000&1&1000&1086&$\bullet$\\
	77&5000&0.05&500&5222&{\color{red}$\bullet$}\\
	78&2000&0.002&200&2000&{\color{green}$\bullet$}\\
	79&10000&0.1&1000&10000\\
	80$^\dagger$&5000&0.5&5000&25000&{\color{blue}$\bullet$}\\
    80a&5000&0.5&5000&6032&\\
	81&5000&0.5&5000&25000&{\color{cyan}$\bullet$}\\
	82$^\dagger$&5000&0.05&500&25000&
	{\color{pink}$\bullet$}\\
	83&2000&20&2000&2971&{\color{yellow}$\bullet$}\\
	84&3000&3 &3000&17000&{\color{brown}$\bullet$}\\
	85&1300&13&1300&1540&{\color{lime}$\bullet$}\\
    \hline
\multicolumn{6}{l}{*Started at z = 1100.}\\
\multicolumn{6}{l}{$^\dagger$Runs 80, 82 used 150000 particles.}\\
\multicolumn{6}{l}{Run 80a used the alternate IMF in Figure 3.}\\
\multicolumn{6}{l}{Bullets are the color code in Figure 4.}\\
\end{tabular}
\end{table}
\begin{figure}
\includegraphics[width=.5\textwidth]{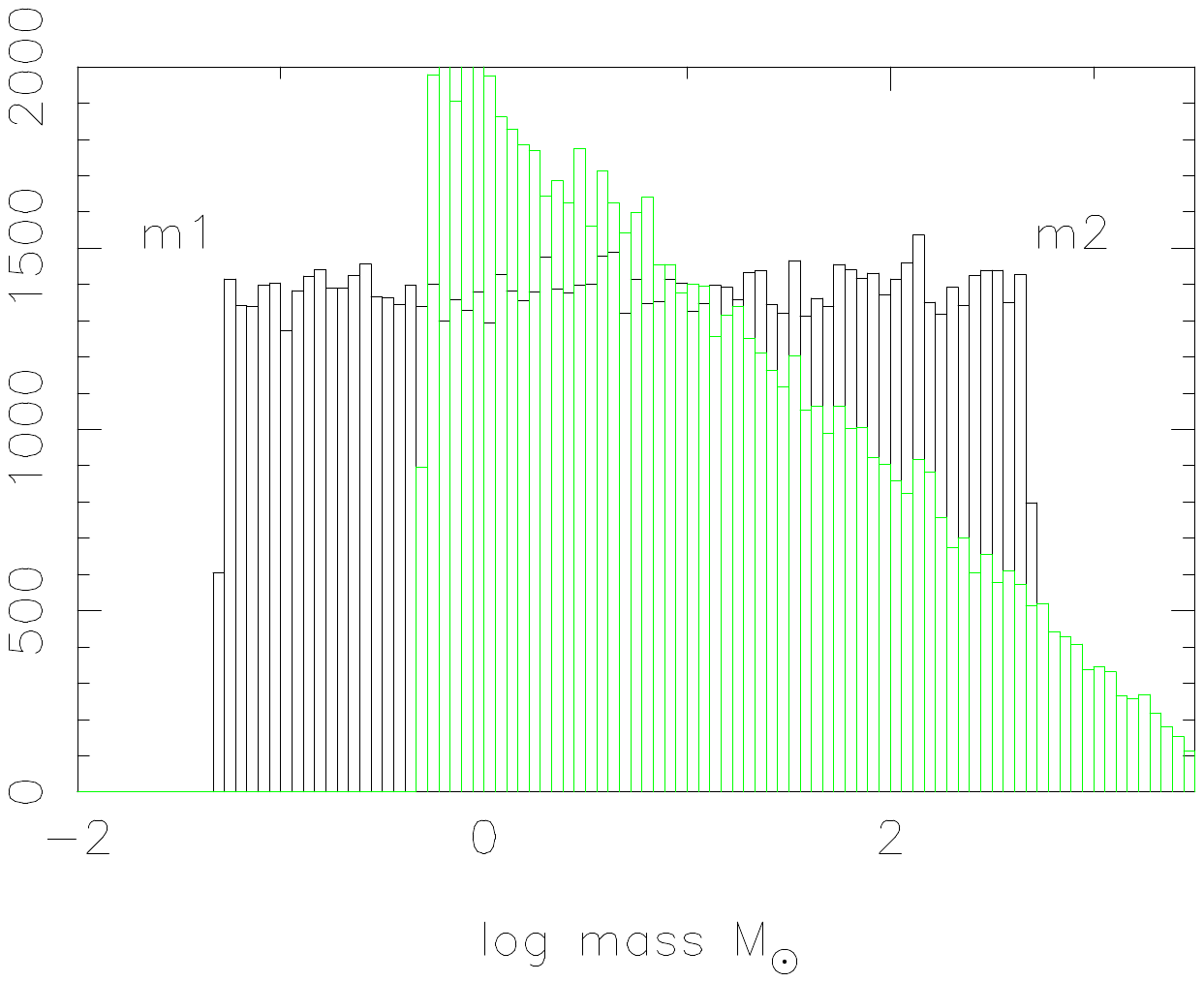}
\caption{The distribution of PBH masses in run 77 (black) and run 80a (green).}
\end{figure}

Figure 4 shows the growth of nuclei in these runs, which we started at z = 1300 close to the microwave background surface of last scattering. Noticeably, accretion is concentrated at early times, before the scale factor has reduced the density.

\begin{figure}	
\includegraphics[width=.66\textwidth]{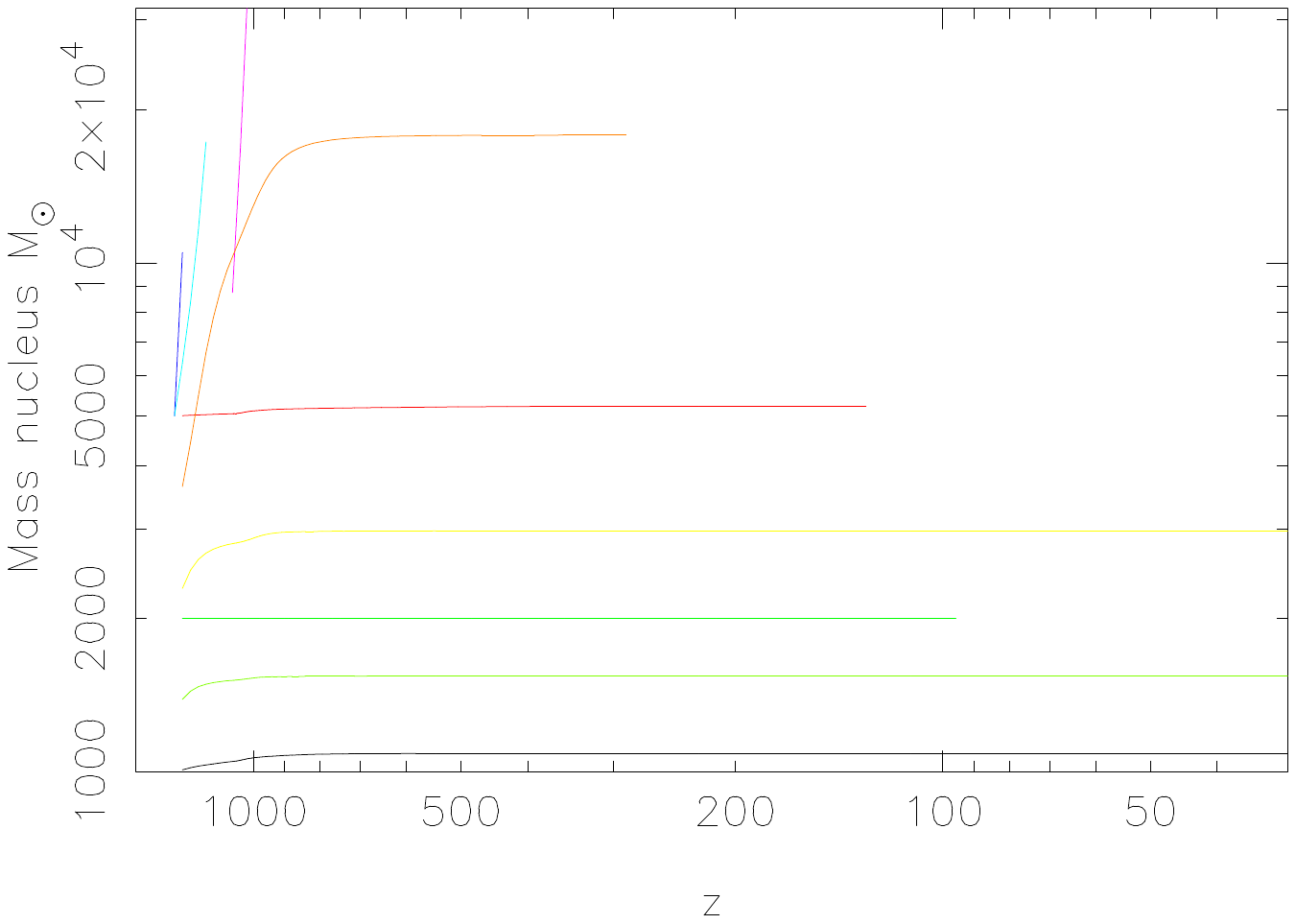}
\caption{Growth of nuclei with redshift by accretion. Colour coding is given in Table 1. For large initial masses the growth is rapid, and the simulation would require a reset to continue. Accretion falls off as the scale factor (1+z)$^{-1}$ increases.}
\end{figure}

\begin{figure}	
\centering
\includegraphics[width=.99\textwidth,angle=0]{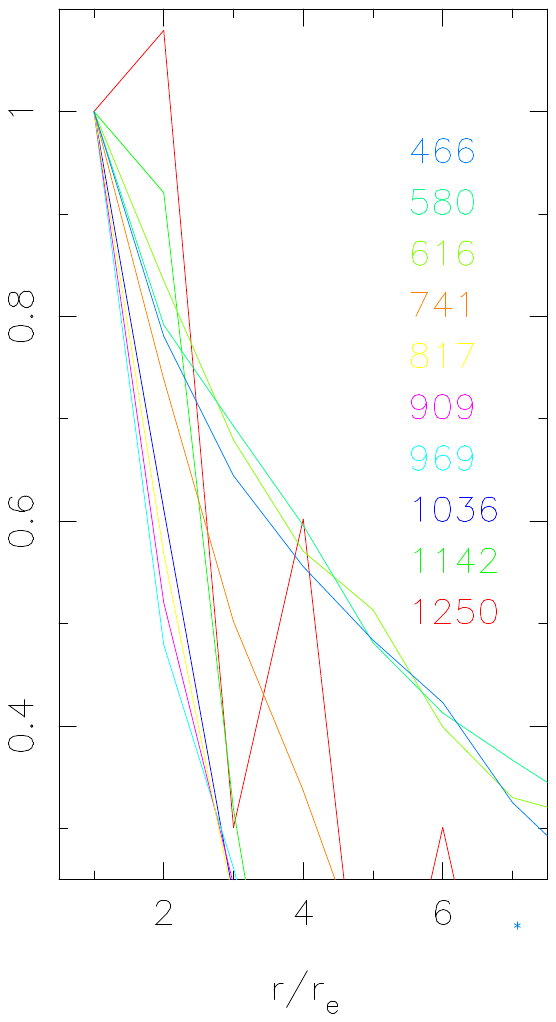}
\caption{Radial profile of mass as a function of redshift in run 84. Redshifts are numbered. The y-axis is the
number of particles per unit projected area normalized to 1; 
the x-axis is radius divided by effective radius. The effective radius
contains half the particles. The structure is initially noisy, contracts towards the nucleus, and then expands.}
\vspace{1 cm}
\end{figure}
\begin{figure}[h]	
\centering
\includegraphics[width=.79\textwidth,angle=0]{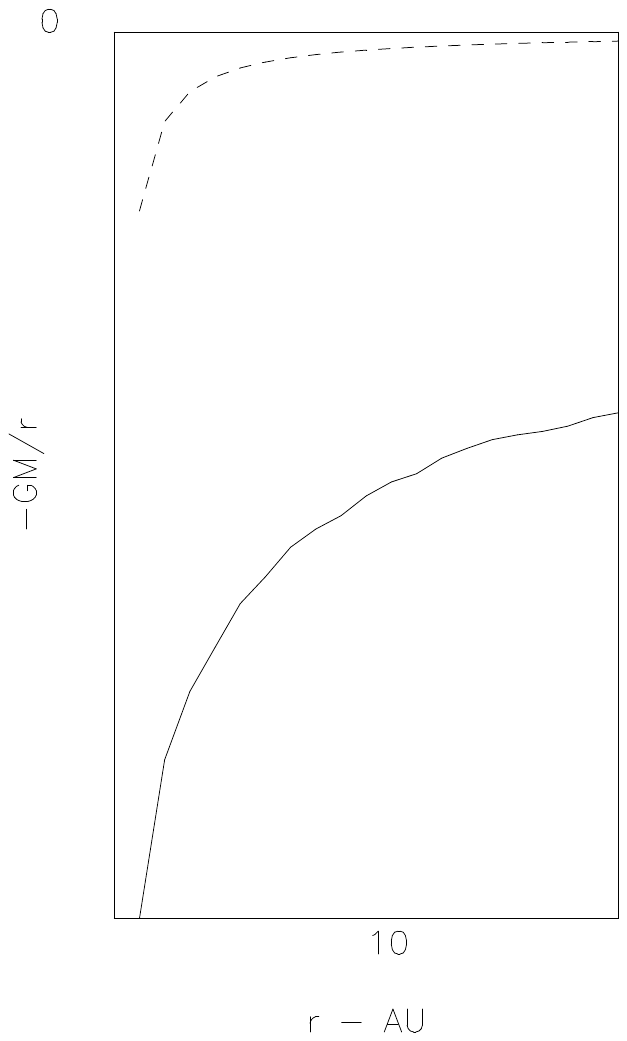}
\caption{Potential well for run 84 at redshift 817 (solid curve). This
has deepened markedly from the original potential of the nucleus alone (dashed curve). The gas and other particles
are drawn into this potential well and, once the gas has cooled, e.g. by redshift 100, and the density is sufficient, formation of a luminous GC can begin.}
\end{figure}
\subsection{Additional parameters}
Besides those trialled in Table 1, we note some parameters still
to be explored. These include power law IMFs beyond n $\sim$ m$^{-1}$
and f$~^\prime$, the fraction of dark matter outside this mass range. 
The 'blue'-tilted IMF
shown as the alternate in Figure~3 decreases the accretion rate.

Although the simulation code is originally scale free, the density dependence of accretion means that units must be adopted in this application. These are astronomical units of solar masses, years and AU. Velocities are then in 30 \kms units. Here we have dealt with the formation of the dark matter potential. In the next section we are concerned with the fate of the gas that falls into the potential (Figure 6) where the density is so much higher than its original state.
\section{Star formation}
The nuclear environment has a strong central concentration (Figure 5). Into this potential well, the gas is assumed to fall. We do not follow this with further simulation, but outline what we expect to happen  to make an observable globular cluster.

Initially the baryons were well mixed with the dark matter, and they
would take part in the accretion into the nuclear region. Run 82 has
a final nuclear PBH mass of 25000 M$_\odot$ and the accompanying baryons $\Omega_b/\Omega_m$ = 15\% of that mass. The central regions of the simulations
have dimensions of 10 AU, and so, without attention to details of the equation of state, densities
$\sim$10$^{-6}$ gm/cm$^3$
are reached, and the Jeans length 
is 0.28$\surd$( kT/eV)  AU for kT $<$ 0.01 eV or T $<$ 116K.   
This would be expected to trigger star formation as soon as z~$\sim$~100. At this time GC sized
gas clouds need little encouragement to collapse (Peebles 1969).

Questions arise about the interaction of the stars and the intermediate black hole (IMBH) nucleus. As in the Milky Way, most of the stars are on orbits that
miss the Schwarzschild radius of the central black hole, which for 10$^4$ M$_\odot$
is 10$^{-4}$ AU. Only an extremely radial fraction come close. Mass loss from
old stars, however, is a steady 0.015 M$_\odot$ yr$^{-1} (10^9$ L$_\odot)^{-1}$
(Faber \& Gallagher 1976). Suppose for a 10$^5$ L/L$_\odot$ cluster, 10\%  of this makes its way to the IMBH nucleus and energy were released at, say, 10\% efficiency, that is 7.6 $\times$ 10$^{4}$ L/L$_\odot$ from the nucleus.
The Hawking radiation temperature is a fraction of a $\mu$K, but if the material formed an accretion disk, that luminosity might emerge in the x-ray and far ultraviolet region (FUV), which is not observed. An old GC's post-AGB
star has been measured at over 2000 L/L$_\odot$ (Kumar \etal 2024), but this is an order of magnitude fainter and not a nuclear object.
Accretion disks feeding compact objects can have number densities from 
10$^{15}$ 
to 10$^{22}$ cm$^{-3}$; so the disk would be unresolved ($<<$ 1 $\mu$as) in the Galaxy. 
\subsection{Relationship to globular clusters}
We have simulated the formation of dark clusters of PBH centered around thousand solar mass dominant objects at high redshift, and stopped short of following the response of
the accompanying baryons to the high densities that result. What happens then ? Our short answer is that pregalactic GCs develop further by accretion. These density
concentrations would draw in additional matter over a free fall time into a gravitational potential of order unity in \kms. Assuming that
massive stars form, we expect supernovae to add metals to the pristine
hydrogen and helium. In initially shallow gravitational potentials
supernovae would readily drive out gas leading to a distribution function of metals orders of magnitude lower in their mean $<Z>$ than $Z_\odot$ 
according to the simple model of chemical enrichment with gas loss (Pagel \& Patchett 1975; Hartwick 1976). 
This is the standard approach to reducing the yield in these and later models (Gibson 2002; Garnett 2002).

This is a scenario for producing blue clusters in galaxies, whose color distribution is well known to be bimodal (Brodie \& Strader 2006). 
Other possible causes of bimodality are worked out by Li \& Gnedin (2014) and Tonini (2013). These involve galaxy mergers. Furthermore, multiple stellar populations, 
which are now known to be common in globular clusters
 (Gratton, Carretta \& Bragaglia 2012), are more likely in this extended history. 
 A pregalactic origin, however, sheds no light on their peculiar chemical abundance distribution. A complementary
supposition would be that red clusters are post galactic: a Gyr or so younger, enriched in
metals, and with red horizontal branches.

\subsection{How pregalactic are the blue clusters?}
Is the term $pregalactic$ globular clusters warranted? The free fall time of baryons is proportional to the reciprocal of $\surd\rho$.
The GCs that form from these "dark" clusters that only
emit Hawking radiation 
are parsec sized major overdensities in the kiloparsec sized
regions that form galaxies. Not only are their dark precursors pregalactic (Figure 1), but they are also pregalactic in their later fully developed form. Such clusters have nuclei that are 13.0
$\pm$ 0.35 Gyrs old\footnote{Mould (2025) suggests that CMB ages may need correction for the effect of a PBH "ionization time-bomb" on another phase transition, recombination. An upper limit on this correction is not difficult (Batten \& Mould 2025), but a lower limit requires assumptions about the PBH IMF.} (Riess \etal 2024). The oldest stars may also be this old within the uncertainties, unless star formation fails to ignite before z = 10, in which case the stars can be 0.46 Gyr younger.  Precision ages of GCs require $monte$-$carlo$ fitting for all of their parameters (van Dyk \etal 2009;
O'Malley, Gilligan \& Chaboyer 2017).

\section{Counter-evidence}
Finding a GC without a 3000 M$_\odot$ nuclear black hole is insufficient to dispose of this proposed mechanism of cluster formation. After the universe is a free-fall time old, there is nothing to stop  10$^6$ M$_\odot$ of baryons from collapsing to a cluster with or without the assistance of dark matter. Finding that, in a large sample of blue GCs, there were no thousand solar mass nuclear black holes, would be compelling evidence against a primordial IMBH formation mechanism, however.
\subsection{X-ray emission }
 Ultraluminous x-ray sources in GCs
are reviewed by
    Wiktorowicz \etal (2025) with luminosities up to and beyond 10$^{39}$ ergs/sec.
    These are extragalactic. Detection of 100 Milky Way GCs in $\gamma$-rays
    by Song \etal (2021) is  interpreted as inverse Compton emission by cosmic-ray electrons and positrons injected by millisecond pulsars. The source positions do not rule out the possibility
    that they are cluster nuclei, however. There is no correlation of luminosity
    and iron abundance from the Harris (1986) catalog. 
Su \etal (2022) conducted a systematic search for a putative IMBH in 81 Milky Way GCs, based on archival $Chandra$ X-ray observations. They found in only six GCs a significant X-ray source positionally coincident with the cluster centre, which have 0.5-8 keV luminosities between $\sim$1 $\times$ 10$^{30}$ erg s$^{-1}$ and $\sim$4 $\times$ 10$^{33}$ erg s$^{-1}. $ However, the spectral and temporal properties of these six sources could also be explained by binary stars. The remaining 75 GCs do not have a detectable central source, most with 3$\sigma$ upper limits significantly lower than predicted for Bondi accretion. From a deep $Chandra$ exposure of $\omega$ Cen Haggard \etal (2013) conclude that either the cluster does not harbour an IMBH, or the IMBH  must experience very little or very inefficient accretion, lower even than the inefficient black hole accreter, Sgr A.
\subsection{Kinematics}
A compilation of black hole detections in GCs was made by L\"{u}tzgendorf \etal (2013). Seven GCs were found to have central black holes $\sim$ 10$^4$ M$_\odot$ with 6 upper limits
$\sim$ 10$^3$ M$_\odot$. The significance of a correlation between black hole mass and [Fe/H] was 0.52.
Fiber optics allow detailed kinematics to be measured in the cores of GCs, e.g. Dalessandro \etal (2021), who studied NGC 6362. If a 1000 M$_\odot$ black hole makes up the cluster's nucleus, the velocity dispersion 2$^{\prime\prime}$ from the center would be 5.4 \kms for a cluster distance\footnote{Circular velocity divided by $\surd$2.} of 7.6 kpc. The cluster's central velocity dispersion is 4.3 $\pm$ 0.4 \kms, which rules out the black hole.
NGC 6362 is not a metal poor cluster, however. Kamann \etal (2014) have studied the archetypal metal poor cluster M92. They set an upper limit on a central IMBH of 980 M$_\odot$ (1$\sigma$) and 2700 M$_\odot$ (3$\sigma$). A second such cluster is M15 where there is a possible black hole detection of 3.9 $\pm$ 2.2 $\times$ 10$^3$
M$_\odot$ (Gerssen \etal 2002). The possibility that neutron stars, rather than an IMBH, raise the central mass to light ratio of GCs should not be ignored (Baumgardt \etal 2003, Hurley 2007).
\subsection{The CMB}
If GCs are $\lesssim$1\% of a galaxy's mass, the spatial power spectrum, on which these ($\sim$~0.1'') point sources are a fluctuation, is of low amplitude. That power is captured by the ln(10$^{10}$A$_s$)~$\approx$~3 parameter in Planck (2020), and so, although dark clusters may be accreting close in time to the CMB photons' last scattering, CMB fluctuations on the finest scales seem unlikely to eliminate or corroborate a pregalactic cluster model. 

\subsection{Intergalactic GCs}
Pregalatic GCs could form anywhere in the model we have described, at least until reionization. If they had formed in the Local Group, they would have been found (Mackey,  Beasley, \&  Leaman  2016).
However, the time for them to fall into the twin large halos of our Galaxy and M31, is measured in Gyrs.  Intergalactic GCs at 10 Mpc are another matter. If present, they would be found in sky surveys at 22nd mag and 1 arcsec resolution. However, our simulations have an initial overdensity which drives the collapse. It is possible that this threshold is not reached in the pregalactic IGM. 
There are, of course, many intergalactic GCs in clusters of galaxies, such as Virgo and Perseus, but the conventional explanation is that they have been stripped from dwarf galaxies  in the clusters.

\subsection{Correlations with galaxy properties }
There is a rich literature on the diversity of GC systems, such as their prevalence in spiral versus elliptical galaxies. Our model would tend to associate these with the red clusters that are made after galaxies have begun to form. 

\section{Conclusions}
\begin{itemize}
\item Recently, the QCD phase transition was noted as a possible environment for the formation of PBHs. 
\item At 1000 M$_\odot$  PBH accretion by dominant mass PBHs can begin, adding more PBHs and accompanying baryonic matter. At lower masses accretion is very slow. 
\item This mechanism predicts PBH nuclei for GCs which would continue to accrete matter at some rate throughout their lives, feeding on some of the mass lost by evolved stars.
\item If an accretion disk is formed, this might be detected in x-rays or the FUV, or it might be obscured by dust and gas.
\item Observations of the oldest Galactic GCs currently do not rule out the presence of nuclear 1000 M$_\odot$ PBHs, but high resolution kinematics of central stars now have the capability to detect them.
\item The formation of globular cluster PBH nuclei, when the universe passes through the QCD transition, may contribute to the gravitational wave background (Vanzan \etal 2024, Hurley \etal 2016).
\end{itemize}

\section*{References}
\noindent
Alonso-Monsalve, E. \& Kaiser, D. 2023, PRL, 132.231402\\
Baade, W. 1944, ApJ, 100, 137\\
Batten, A. \& Mould, J. 2025, submitted to MNRAS\\
Beasley, M. \etal 2003, ApJ, 596, L187\\ 
Bicknell, G. \& Henriksen, R. 1979, ApJ, 232, 670\\
Bird, S., Flynn, C., Harris, W. \& Valtonen, M. 2013, AAS, 221, 30301\\
Bondi, H. \& Hoyle, F. 1944, MNRAS, 104, 273\\
Brodie, J. \& Strader, 2006, ARAA, 44, 193\\
Baumgardt, H. \etal 2003, ApJ, 582, L21\\
Carr, B. 1981, MNRAS, 194, 639\\
Carr, B., Clesse, S., Garcia-Bellido, \& Kuhnel, F. 2019,
arxiv 1906.08217//
Carr, B. \& Kuhnel, F. 2021, arxiv 21100282\\
Cen, R. 2001, ApJ, 560, 592\\
Chen, Z-C. \& Hall, A. 2024 arXiv 2402.03934\\
 Davies, M., Miller, M. \& Bellovary, J., 2011, ApJL, 740, L42\\
Faber, S. \& Gallagher, J. 1976, ApJ, 204, 365\\
Fall, M. \& Rees, M. 1985, ApJ, 298, 18\\
Forbes, D. \etal 2018, RSPSA, 474, 20170616\\
Garnett, D. 2002 ApJ, 581, 1019\\    
Gerssen, J. \etal 2002, AJ, 124, 3270\\
Gibson, B. 2002, ASPC, 273, 205    \\
 Gratton, R., Carretta, E. \& Bragaglia, A. 2012, A\&ARv, 20, 50\\
Haggard, D. \etal 2013, ApJL, 773, L31\\
Harris, W.  1996, AJ, 112, 1487\\ 
Hartwick, D. 1976, ApJ, 209, 418\\
Haselgrove, B. \& Hoyle, F. 1956, MNRAS, 116, 527\\
Hawking, S., 1971,  MNRAS, 152, 75\\
Hurley \etal 2016, PASA, 33, 36\\
Hurley, J. 2007, MNRAS, 379, 93\\
Jedamzik, K. 2025 in {\it Primordial Black Holes}, eds: Byrne, C. \etal Springer Singapore\\
Kamann, S. \etal 2014, A\&A, 566, A58\\
Kruijssen D., J. 2025, arxiv 2501.16438\\
Kumar, R. \etal 2024, A\&A, 685, L6\\
Li, H. \& Gnedin, O. 2014, ApJ, 796, 10\\
Lupi, A. \etal 2014, MNRAS, 442, 3616\\
L\"{u}tzgendorf, N. \etal 2013, A\&A, 555, A26\\
Mackey, A.,  Beasley, M. \&  Leaman, R. 2016, MNRAS, 460, L114 \\
Moore, B. \etal 2006, MNRAS, 368, 563\\
Mould, J.  2025, ApJ, 984, 59\\
Mould, J. \& Batten, A. 2025,  arxiv 2507.11023\\
Musci, I, Jedamzik, K. \& Young, S. 2023, arxiv 2303.07980 \\
Niikura, H. \etal 2019, Nature Astronomy, 3, 524\\
O'Malley, E., Gilligan, C. \& Chaboyer, B. 2017, ApJ, 838, 162 \\
Padoan, O., Jimenez, R. \& Jones, B. 1997, MNRAS, 285, 711\\ 
Pagel, B. \& Patchett, B.1975, MNRAS, 173, 13\\
Peebles, P. J. 1969, ApJ, 157,1075\\
Peebles, P.,J. \& Dicke, R. 1968, ApJ, 154, 891\\
Planck collaboration 2020, A\&A, 641, A6\\
Postnov, K. \& Chekh, I. 2024, arxiv 2407.16273\\
Renaud, F., Bournaud, F. \& Duc, P. 2015, MNRAS. 446, 2038\\
Renzini, A. 2017, MNRAS, 469, 63\\
Riess, A. \etal 2024, ApJ, 977, 120\\ 
Sandage, A. 1953, PhD thesis, Caltech\\
Senchyna, P., Plat, A., \& Stark, D. 2024, MNRAS, 529, 3301\\
Sobrinho, J. \& Augusto, P. 2024, MNRAS, 531, L40\\
Song, D. \etal 2021, MNRAS, 507, 5161\\
Su, Z. \etal 2022, MNRAS, 516, 1788\\
Thomas, D., Kopp, M. \& Skordis, C. 2016, ApJ, 830, 155\\
Tonini, C. 2012, ApJ, 762, 39\\
Trenti, M., Padoan, O. \& Jimenez, R. 2015, ApJ, 808, 35\\
van Dyk, D. \etal 2009, Annals Applied Stats, 3, 117\\
Vanzan, E. \etal 2024, JCAP, 10, 014\\ 
Vanzella, E. \etal 2022, ApJ, 940, L53\\
Volonteri, H., Habouzit, M. \& Colpi, M. 2021, Nature Reviews Physics, 3, 732\\
Wiktorowicz, G.,  \etal 2025, A\&A, 696, 90\\
Wilson, C., Harris, W.  Longden, R. \& Scoville, N. 2006 ApJ, 641, 763\\
Ziparo, F., Gallerani, S. \& Ferrara, A. 2025, JCAP, 4, 40\\ 

\subsection*{Acknowledgements}
The ARC Centre of Excellence for Dark Matter Particle Physics is funded by the Australian Research Council. Grant CE200100008. We thank Duncan Forbes and Jean Brodie for sharing their knowledge of GCs, and the Centre's Nicole Bell for suggesting the QCD transition be put on PBH evolutionary tracks. Simulations were carried out on 
Swinburne University's OzSTAR \& Ngarrgu Tindebeek supercomputers, the latter named by Wurundjeri elders and translating as "Knowledge of the Void" in the local Woiwurrung language. 
\subsection*{Code availability}
There are many available codes that integrate m$_i$d{\bf v}$_{i,j}/dt = Gm_im_j${\bf r}$_{ij}/r^3_{ij}$. We thank Robin Humble for help parallelizing ours\footnote{https://www.intel.com/content/www/us/en/developer/tools/oneapi/base-toolkit-download.html}. It
is available at github/jrmould/darkmatter and may be useful as a parallelization demonstration. See Mould (2025) for the code for Figure 1.
\end{document}